\documentclass[sigconf,nonacm]{acmart}
\acmConference[ISSTA 2023]{ACM SIGSOFT International Symposium on Software Testing and Analysis}{17-21 July, 2023}{Seattle, USA}

\usepackage[T1]{fontenc}

\usepackage{tikz}
\usetikzlibrary{arrows.meta}
\usepackage{tikz-cd} 

\usepackage{subcaption}

\usepackage{enumitem}

\usepackage{isabelle,isabellesym}
\isabellestyle{it}

\usepackage{paper}

\makeatletter
\AtBeginDocument{%
  \def\doi#1{\url{https://doi.org/#1}}}
\makeatother

\makeatletter
\def\comp@sym{\raise 0.6ex\hbox{\small\oalign{\hfil%
        $\scriptscriptstyle\mathrm{o}$\hfil%
        \cr\hfil$\scriptscriptstyle\mathrm{9}$\hfil}}}
\newcommand{\semi}{\mathrel{\comp@sym}}
\makeatother

\newcommand{\Node}[1]{\textsf{#1Node}}

\definecolor{greenhighlight}{RGB}{184,221,209}
\newcommand{\highlightnode}[1]{\colorbox{greenhighlight}{#1}}

\title[Differential Testing of Optimizations]{Differential Testing of a Verification Framework \texorpdfstring{\\}{}for Compiler Optimizations (Experience Paper)}

\author{Mark Utting}
\orcid{0000-0003-3134-6306}
\affiliation{
  \institution{The University of Queensland}
  \country{Australia}
}
\email{m.utting@uq.edu.au}

\author{Brae J. Webb}
\orcid{0000-0003-3579-0244}
\affiliation{
  \institution{The University of Queensland}
  \country{Australia}
}
\email{b.webb@uq.edu.au}

\author{Ian J. Hayes}
\orcid{0000-0003-3649-392X}
\affiliation{
  \institution{The University of Queensland}
  \country{Australia}
}
\email{ian.hayes@uq.edu.au}



\begin{document}

\begin{abstract}
We want to verify the correctness of optimization phases in the GraalVM compiler, which consist of many thousands of lines of complex Java code performing sophisticated graph transformations.
We have built high-level models of the data structures and operations of the code using the Isabelle/HOL theorem prover, and can formally verify the correctness of those high-level operations.  But the remaining challenge is: how can we be sure that those high-level operations accurately reflect what the Java is doing?  This paper addresses that issue by applying several different kinds of differential testing to validate that the formal model and the Java code have the same semantics. 
Many of these validation techniques should be applicable to other projects that are building formal models of real-world code.
\end{abstract}

\maketitle

\section{Introduction}

GraalVM \cite{graal} is a widely-used, high-performance, polyglot compiler that supports Java and many other languages, so the correctness of its compiler is of high importance.
There are several front ends to the compiler,
including JVM bytecode for Java and other JVM languages,
JavaScript bytecode for running JavaScript and Node.js programs,
LLVM bitcode for executing C/C++, Rust, Julia, etc.,
and Truffle interpreters for many other languages.
The compiler can be used for hotspot compilation, where heavily used methods are compiled and optimized during execution,
or for native compilation to generate an optimized executable binary. 

We are formally verifying optimization transformations used in the GraalVM compiler by building formal models of the compiler's Intermediate Representation (IR) \cite{Stadler2013GraalIR} and its execution semantics~\cite{ATVA21_GraalVM_IR_Semantics,sea-of-nodes-semantics}, defining optimization transformations, and verifying that they preserve the execution semantics.  Most of these optimizations are common between the hotspot and native image modes of the compiler.  We are using the Isabelle/HOL theorem prover for our models and verification, to obtain the highest level of assurance that the verified optimizations are correct according to our IR semantics.

However, as with all formal verification, a possible weak link is the connection between the formal models and the real world -- the thousands of lines of Java code that implement the IR and its high-level optimizations (the GraalVM compiler also includes many low-level machine-dependent optimizations, but they are outside the scope of this paper).

This paper addresses two research questions relating to validation issues between formal models and the real world:
\begin{enumerate}
  \item How can we validate that our IR semantics in Isabelle matches the expected semantics of the GraalVM compiler IR?\\
  This is non-trivial because the compiler IR has no formal semantics and its nodes do not always directly correspond with JVM constructs because the IR has to be sufficiently general to support all its hosted languages.

  \item How can we be sure that our formal descriptions of each IR optimization \emph{transformation} correctly match the transformations that are implemented by Java code in the compiler?\\
  This is challenging to ensure in general because many optimization transformations have complex pre-conditions that determine whether the optimization is valid, and in the Java code of the compiler these pre-conditions are typically spread out over hundreds of lines of code and multiple Java classes because they may apply to many related optimizations.  
\end{enumerate}
Our validation of the Isabelle/HOL IR semantics is split into two phases:
validation techniques for individual arithmetic operators (see Section \ref{sec:arith}), 
and 
validation of our complete IR Isabelle semantics by automatically translating the existing GraalVM compiler unit tests into our Isabelle IR notation and 
using them to validate the control-flow and data-flow semantics (see Section \ref{sec:execsemantics}).
In Section \ref{sec:iropt} we describe how we validate our Isabelle definitions of IR optimization transformations against the actual transformations in the compiler, by differential testing \cite{McKeeman1998DifferentialTesting} of the Isabelle transformations against the Java transformations.
Our approach leverages the numerous existing (Java) test cases already used to test the GraalVM compiler.

\section{Validating fixed-width arithmetic} \label{sec:arith}

The GraalVM IR is a \emph{sea of nodes} data structure \cite{Click1995,click95,duboscq:ir:2013} that combines the control-flow graph and the data-flow expression graphs into a single graph structure, with hundreds of different kinds of nodes.
The semantics of expressions has to correctly handle all the different data types, such as fixed-width integers of 1, 8, 16, 32, and 64 bits (signed and unsigned in some cases), to accurately implement the semantics of all the different languages supported by GraalVM.  Our Isabelle semantics supports all data types except for floating-point, which is left for future work.

Our first step towards validating our expression semantics is to systematically test the semantic definition of each integer operator, using the boundary values of each fixed-width integer type.
Isabelle includes a $word$ library that defines fixed-width integers of arbitrary bit-width $b$ by the type, $b~word$, and defines a comprehensive set of operators and theorems useful for proving properties of those operators.  We use this library to implement all the GraalVM expression operators, taking care to correctly handle the different integer widths.  

\paragraph{Testing Rationale.}
One of our key motivations for wanting good systematic testing of the integer operators was that our Isabelle/HOL representation of Java data values changed over time, as we searched for the best representation that captures the Java semantics accurately but also enables Isabelle/HOL to easily prove results about integer operators.
Each change of our representation required consequent changes to our Isabelle/HOL definitions of the Java arithmetic operators, so it was important to use the tests as regression tests to find any errors in our operator definitions.

We initially used the unbounded Isabelle $int$ type to represent all integer results, but as Java does all integer calculations with either 32 or 64-bit integers, we changed to represent these as two distinct data types in our semantics:
\begin{eqnarray*}
datatype && Value =\\
 &&  UndefVal\ |\\
 &&  IntVal32\ (32\, word)\ | \\
 &&  IntVal64\ (64\, word)\ | \ldots
\end{eqnarray*}
This made it easy to use the existing Isabelle $word$ library operators with the correct modulo behavior for each integer width.  However, over time we found that this representation required too much case analysis in our correctness proofs, and it also made it more difficult to define the IR narrowing and widening (both zero-extend and sign-extend) operators, which change the bit width and manipulate a larger set of integer widths ($1,8,16,32,64$ bits).  Consequently we switched to a unified representation of integers that records the explicit number of bits that are significant and always uses a 64-bit word for the value (with unused upper bits being zero):
\begin{eqnarray*}
  datatype && Value =\\
   &&  UndefVal\ |\\
   &&  IntVal\ nat\ (64\, word)\ | \ldots
\end{eqnarray*}
This simplified many proofs, and simplified the definitions of the narrowing and widening operators, but made some other arithmetic operator definitions slightly more complex because they now have to explicitly manage the modulo arithmetic of the required number of bits.  

These changes of representation and operator definitions made it crucial to thoroughly test that the semantics of our IR arithmetic operators in Isabelle correctly implemented the corresponding JVM operators used in the GraalVM IR.  From the beginning of the project we relied on automated test generation to test our Isabelle operator definitions against the corresponding Java operators.  

\paragraph{Test Generation.}
To test each operator systematically, we wrote a simple test generator in Java that generated a small IR graph corresponding to a method to test each operator, and then executed that method using our executable IR semantics, iterating over a set of boundary values for each input parameter.
For example, for the Java left shift operator `<{}<' it generates the IR graph shown in Fig.~\ref{fig:leftshiftGraph}, which corresponds to the following Java method:

\begin{lstlisting}[language=Java,frame=none,linewidth=0.92\columnwidth,xleftmargin=0.7cm,framexleftmargin=4pt]
static int leftShiftNode32(int a, int b) {
  return a << b;
}
\end{lstlisting}

\begin{figure}[ht]
\includegraphics[width=0.2\textwidth]{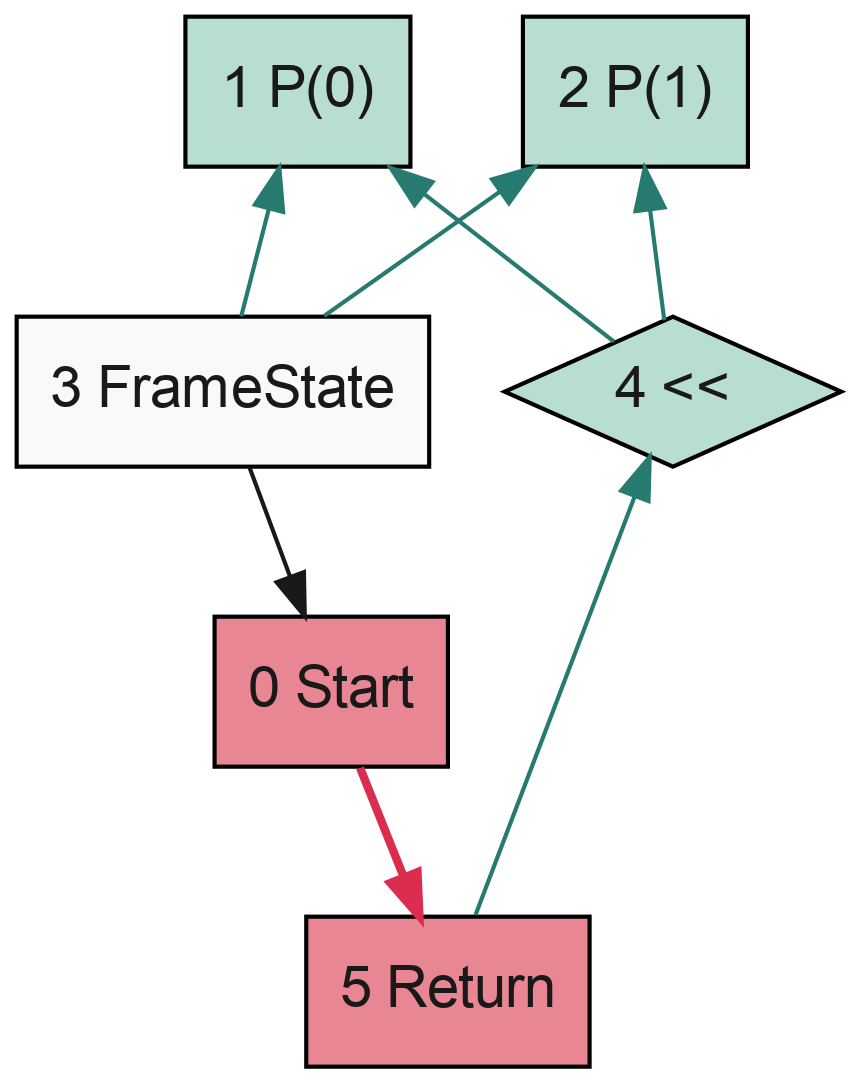}
\caption{Generated IR graph for the left shift method.}
\label{fig:leftshiftGraph}
\end{figure}

Each method has a \Node{Start} with $id=0$, and execution starts at this node and follows the \emph{successor} edges of the control-flow graph (0 then 5 in this case),
and calculates expression values when needed by using backward \emph{data-flow} edges to find the inputs of each operator.
Nodes 1 and 2 represent parameters \textsf{P(0)} and \textsf{P(1)}, i.e.\ $a$ and $b$, respectively.
When the \Node{Return} is reached, it evaluates its expression (node $4$), that results in $a << b$ and this result returned to the caller.
The \textsf{FrameState} nodes are not used by our Isabelle semantics but are used in the Java GraalVM IR to reconstruct stack frames when deoptimizing if speculative optimization fails \cite{duboscq:ir:2013}.

\begin{figure*}
\begin{lstlisting}[frame=none,linewidth=0.8\textwidth,xleftmargin=4cm,framexleftmargin=4pt]
definition leftShiftNode32 :: IRGraph where
  "leftShiftNode32 = irgraph [
    (0, (StartNode (Some 3) 5), VoidStamp),
    (1, (ParameterNode 0), IntegerStamp 32 (-2147483648) (2147483647)),
    (2, (ParameterNode 1), IntegerStamp 32 (-2147483648) (2147483647)),
    (3, (FrameState [] None None None), IllegalStamp),
    (4, (LeftShiftNode 1 2), IntegerStamp 32 (-2147483648) (2147483647)),
    (5, (ReturnNode (Some 4) None), IntegerStamp 32 (-2147483648) (2147483647))
   ]"
lemma "static_test leftShiftNode32
    [(IntVal32 (2)), (IntVal32 (2))]
    (IntVal32 (8))" by eval
lemma "static_test leftShiftNode32
    [(IntVal32 (1)), (IntVal32 (2))]
    (IntVal32 (4))" by eval
lemma "static_test leftShiftNode32
    [(IntVal32 (0)), (IntVal32 (2))]
    (IntVal32 (0))" by eval
...
\end{lstlisting}
\caption{Generated IR graph for the left shift operator, plus first few tests.}
\label{fig:leftshift}
\end{figure*}

Our Isabelle representation of IR graphs is a mapping from node identifiers (natural numbers) to a pair consisting of a node and its stamp.
Fig.~\ref{fig:leftshift} shows the Isabelle representation of the IR graph for the left shift test method in Fig.~\ref{fig:leftshiftGraph}, 
and the first few tests that we generated for this operator.
The nodes are similar to the leaf classes of the Java IR Node class hierarchy, and the stamps give static typing information for each node.
Note that the Isabelle $option$ type is used to represent optional values, so $None$ corresponds to a missing value (typically a null pointer in the Java IR),
and, $Some~n$, a (existing) value of $n$.
Each integer expression node is paired with an $IntegerStamp$ that describes the number of bits in the integer result of evaluating that node according to its type, plus statically determined minimum and maximum integer bounds on the integer that may be returned. The stamp $default\_integer\_stamp$ is an abbreviation of $IntegerStamp~32~MIN~MAX$, where $MIN = -2147483648$ and $MAX = 2147483647$.

We generate this style of IR method graph for each IR operator, and test it using the test harness $static\_test\ m\ i\ o$, where $m$ is the method graph, $i$ is the list of input parameter values, and $o$ is the expected output value.  We obtain the expected output value $o$ by executing the program in Java, with inputs $i$.  This ensures that we are testing our Isabelle IR semantics of each operator against the Java/JVM semantics of that same operator.

We test the 32-bit and 64-bit versions of each operator separately, because the JVM typically has separate bytecode operators for these.  For each input parameter, we iterate over a set of boundary values near the $MIN$ and $MAX$ values of that parameter's type, plus values around zero:
\[
  values = \{MIN, MIN+1, -2,1,0,1,2, MAX-1, MAX\}
\]

However, for the division and modulo operators we exclude $0$ as the second parameter because exceptions are handled via different mechanisms in the IR.
This generates 81 test cases for most binary operators and 9 for each unary operator.
The Isabelle tactic $eval$ uses simple evaluation rules derived from the operator definitions to evaluate expressions.  This tactic will fail if the evaluated output does not equal the expected output value, and such failures are flagged in Isabelle.

By running these generated tests as we developed the corresponding Isabelle definitions of the 32-bit and 64-bit integer operators, we uncovered some interesting bugs in our definitions, such as:
\begin{itemize}
  \item the $word$ library $div$ operator is unsigned, whereas Java requires signed division, so we must use the $sdiv$ from the $Signed\_Division$ library (similarly for $mod$).

  \item due to the twos-complement representation, the maximum negative value often has unusual behaviour.  For example, when using unbounded integers we found one failing test was $-2147483648 / -1 == -2147483648$ for 32-bit division, and a similar case for 64-bit division, because this is the only case where division can overflow.

  \item the XOR operator initially failed some tests because unused bits in the output became non-zero.
  The fix was to mask bits outside the bit width of the type of the result to ensure they were zero. 

  \item the \Node{SignedDiv} and \Node{SignedRem} are different to other integer operators because they may throw divide-by-zero exceptions, so they have to be executed as control-flow nodes before they can be evaluated within data-flow expressions.  This would automatically be the case when translating existing GraalVM IR graphs into Isabelle/HOL, but our simple test generator was generating Isabelle/HOL IR graphs directly, so initially treated these two operators the same as other integer operators and did not add them to the control-flow path, which caused the tests to fail.  This is an example of a bug in the test generator itself, rather than the semantics that is being tested.

  \item we had to change \Node{RightShift} to insert sign bits at the correct position in the word (bit $b$), then forgot to handle the positive case where no sign bits are propagated!  The arithmetic tests for this operator detected this quickly.
\end{itemize}

\section{Validating our IR semantics} \label{sec:execsemantics}

The GraalVM compiler has thousands of unit tests, to thoroughly test the optimizations that the compiler provides.  We want to take advantage of that large suite of carefully designed tests, by using it to help validate our IR execution semantics.

Many of those unit tests work by passing a method name and some input parameters to a $test(m,args...)$ helper method, which \emph{compares} the outputs of the unoptimized and optimized versions of the method, to make sure that optimization preserves the semantics of the method.  More precisely, this helper method first compiles and executes the method without optimization and stores the result value $V$, then compiles the method with optimization, executes it, and checks that the new result equals $V$.

We utilize these tests by adding a third comparison, to check that our Isabelle IR semantics also returns the same result.  To achieve this, we:
\begin{enumerate}
  \item wrote a translator from the Java GraalVM IR graph to our Isabelle representation of the IR graph;
  \item modified $test$ helper method that captures the IR graph of the method, the input parameter values, and the expected result value from the unoptimized Java execution, and writes all these out to an Isabelle theory file; and
  \item ran Isabelle on the generated theory file to define each method graph, execute each test case with the appropriate inputs and the expected result value.
\end{enumerate}
Some additional pragmatic points that make this process practical are:
\begin{itemize}
  \item the translation of the Java unit tests into Isabelle only needs to be done occasionally when the compiler changes the semantics of its IR, which is not frequent (note that we are translating the un-optimized IR graph, so the generated tests are independent of any optimizations that the compiler performs - they depend only upon the IR semantics), and we store the generated Isabelle tests and can run them at any time, independently from the execution of the Java unit tests;
  \item the IR graph translator throws an exception when it finds any IR nodes that are not supported by our Isabelle semantics, as we currently implement only around 50 of the most-used high-level IR nodes.  Our modifications to the $test$ helper method skip any tests whose method graph throws a translation exception, so that all translated tests are in the IR subset that we support in Isabelle.
  \item the GraalVM compiler \texttt{unittest} command has options to run any subset of the unit tests, so we can use these to generate a suite of Isabelle unit tests focussed on a particular language feature.  We also added an option to filter out any method tests that rely on floating point inputs or outputs, as our Isabelle semantics does not yet implement floating point (and the same IR nodes are used for floating point and integer arithmetic).
\end{itemize}

By running these translated unit tests, we found some interesting bugs and omissions in our Isabelle IR semantics, plus a few omissions in our IR translator:
\begin{description}
    \item[Static fields:] Many of the methods being tested relied on static field values in their class (or even in other classes) 
    but these static fields were not being translated into the Isabelle and the Isabelle environment did not support their initialisation, 
    so we extended our translator to capture the value of those static fields and 
    include initialization code in the generated Isabelle IR graphs to recreate those static fields with their initial values.

    \item[Boolean versus int:] In the JVM, Boolean values are represented by integer values 0 and 1.  However, in the GraalVM IR these `Boolean' values may be represented as 1-bit integers or 32-bit integers during calculations, depending upon how the value was calculated.  A large percentage (97\%) of our translated test methods with Boolean results initially failed, simply because of mismatches between the expected and actual bit size of the result values.  To fix this we modified our IR graph translator to coerce Boolean results to 32-bit integer values.
    \item[Helper methods:] Many of the methods being tested called other methods, either Java library methods or user-defined methods.  So we extended our translator to translate all (directly or indirectly) called methods into Isabelle, thus generating an `IR Program' in Isabelle, which maps each method name to its IR graph.

    \item[Function results:] When we changed our representation of expressions, some methods containing function calls failed.  This was because function calls were being executed correctly but their return value were not passed into the expression evaluation environment.

    \item[Signed comparisons:] Our integer $<$ operator was using the Isabelle word $<$ operator, but that was unsigned, as pointed out by several failing tests.  We needed to use signed integer comparison $<_s$ from the $word$ library.

    \item[Unicode:] Some tests used constant strings with unusual Unicode characters.
  The Isabelle/HOL parser does not support full Unicode (even though the JEdit user interface does), so we decided to throw an exception and skip tests with Unicode characters that cannot be translated into Isabelle strings.  Eventually, we may implement our own Java-like Unicode strings within our Isabelle semantics, but that is not a current priority.

\item[Dynamic generation:] One test (\textsf{SubWordReturnTest}) generated classes and static fields dynamically, so those static fields were not detected by the our translator.  We decided to skip such tests, so added that test to a `do not translate' list.

\item[Performance:] As we translated more unit tests into a single theory file (37 thousand lines), Isabelle execution slowed down significantly.  So we split the output into multiple files of around 1000 lines each.  Then all files executed reasonably quickly except for one, which had a method containing a loop of 100000 iterations. 
  That test passed if the loop iteration constant was reduced to 100, so we ignored that test, as its long execution time was purely a performance issue, and performance is not relevant for our Isabelle semantics.

\item[Numeric promotion:] Some shift operator tests failed, because our Isabelle semantics was incorrectly doing binary numeric promotion of argument types, instead of unary promotion \cite[Chapter 5.6]{JLS17}. 
  Shift operators are different to the other Java binary operators in that the two inputs can be different sized integers and the type of the result depends only on the type of the left input.

\item[Dynamic dispatch:] Some tests fail because they call methods dynamically, using dynamic dispatch.
  Our Isabelle semantics does not handle this yet, due to a lack of class hierarchy information (in future we plan to also translate class typing information and subclass relationships).

\item[GraalVM-specific compiler directives:] One of the \emph{deoptimization} tests (\textsf{DeoptimizeDirectiveTest}) failed because it used GraalVM-specific compiler directives to detect whether it was running in compiled or interpreted code.  We filtered this test as being out of scope.

\item[Concurrency:] One test (\textsf{Thread\_yield01}) failed because it relies on a JUnit \verb!@Rule! annotation to generate a timeout that interrupts the test.  Threading is out of scope for our Isabelle semantics at the moment, so we ignore this test.

\item[Small integer parameters:] test method input parameters of type short, char, and byte had to be expanded into 32 bit integer, to match Java semantics.  This was another case where the bug was in our IR translator, rather than our Isabelle/HOL IR semantics.

\item[64 versus 32-bit comparison:] A minor error in our definition of signed less than ($<_s$) meant that \Node{IntegerLessThan} and \Node{Abs} were comparing the whole 64 bits rather than extracting (signed) number of bits specified by the type of the operand.

\item[\Node{SignedDiv}:] This had an off-by-one error in extracting the signed value 
    (the Isabelle $signed\_take\_bit$ function has an unusual API that is different to the $take\_bit$ function).

\end{description}

These examples illustrate how effective the translated GraalVM compiler unit tests were in detecting overt and subtle bugs in our IR semantics.
After fixing these issues our Isabelle IR semantics now passes 2538 of the 2544 translated unit tests (99.76\%), and fails just 6 of the tests (0.24\%).  
The 6 failing tests are related to the handling of dynamic method calls in objects whose type is unknown at compile time.  Our IR translator currently generates an IR with a statically-resolved \Node{CallTarget}, which is not correct for dynamic dispatch cases.  We plan to handle dynamic dispatch in the future, but the Isabelle model currently has no information about class structures or hierarchies, because that information is not included in the IR graph.

\section{Validating optimization transformations} \label{sec:iropt}

In this section we turn to the second question posed: \emph{How can we be sure that our Isabelle descriptions of optimization transformations correctly match the transformations actually implemented in the compiler?}

\begin{figure}[H]
\[ \begin{tikzcd}[column sep=small]
        && g' \arrow{rr}{E} && i'' \arrow[leftrightarrow]{dd}{\approx_s} \\
    g \arrow{urr}{G} \arrow{drr}{E} && \\
        && i \arrow{rr}{I} && i'
\end{tikzcd}
\]
\caption{Process overview for validating optimization specifications}
\label{fig:commuting}
\end{figure}
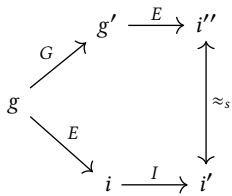

We approach validating the faithful specification of optimizations using
differential testing as illustrated by Figure \ref{fig:commuting}.

The meaning of the nodes in this diagram is as follows:
\begin{description}[leftmargin=2cm,labelsep=2em,align=left,labelwidth=1.5em,itemindent=0cm]    
    \item[$g$] The initial program in GraalVM IR.
    \item[$g'$] The graph of the program in GraalVM IR after applying optimizations in the GraalVM compiler.
    \item[$i$] $g$ encoded in our Isabelle/HOL representation.
    \item[$i'$] The program, in our Isabelle/HOL representation after applying optimizations expressed in Isabelle/HOL.
    \item[$i''$] $g'$ encoded in our Isabelle/HOL representation.
    \item[$G$] A transition representing a GraalVM optimization.
    \item[$I$] A transition representing an Isabelle/HOL optimization.
    \item[$E$] A transition representing encoding a GraalVM graph structure in our Isabelle/HOL representation.
    \item[$\approx_s$] Structural equivalence ignoring specific node identifiers.
\end{description}

We begin with a GraalVM IR graph $g$ for a program,
That graph is then optimized in GraalVM to produce an optimized graph, $g'$.
Graphs $g$ and $g'$ are both then  encoded into the Isabelle/HOL representation to produce graphs $i$ and $i''$ respectively.
The unoptimized graph, $i$, is optimized using the Isabelle/HOL optimization specification, $I$, to produce $i'$.
Finally, the GraalVM optimized graph, $i''$,
is structurally compared ($\approx_s$) to the Isabelle optimized graph, $i'$, 
where the structural comparison allows for different node identifiers in the two graphs, but checks that the structure is equivalent.

So we are differentially testing the $g \longrightarrow g'$ optimization transformation, as implemented by Java code in GraalVM compiler, 
against the corresponding $i \longrightarrow i'$ optimization transformation functions, which are defined within our Isabelle models.
This can be captured by the following equivalence in which $\semi$ is forward composition of transformations.
\begin{equation}\label{commute}
  G \semi E \approx_s E \semi I
\end{equation}

For a collection of programs,
we optimize each program in GraalVM and in Isabelle/HOL using our optimization specification.
If the optimized programs are structurally equivalent,
then we have some assurance that the specification is faithful.

When generating optimization validation tests,
we utilize the same \texttt{test(m, args...)} helper for validing execution semantics.
A graph is constructed for the method corresponding to the method name, $m$.
The constructed graph is used as the initial graph, $g$.
Optimizations are then applied to the initial graph to produce the optimized graph, $g'$.
Figure \ref{fig:example1} presents an example of this process, illustrating Java programs that correspond to the IR graphs.
Optimization of conditionals is split into two phases:
\begin{itemize}
\item
a conditional elimination phase that replaces the condition within an \Node{If} with \textsf{true} if the condition is known to hold from the context and
with \textsf{false} if it is known to not hold---in the example this is the transformation from Fig.~\ref{fig:example1unopt} to Fig.~\ref{fig:example1opt}; and
\item
as part of the canonicalization phase, 
replacing a conditional that has a constant, \textsf{true} or \textsf{false}, condition with its ``then'' or ``else'' branch, 
respectively---this is the transformation from Fig.~\ref{fig:example1opt} to Fig.~\ref{fig:example1opt2}.
\end{itemize}
Our approach differentially tests each of these phases separately, making it easier to isolate any failures to a particular phase.
It is desirable to be able to test optimization phases independently to allow incremental development of optimization specifications.
The canonicalization phase, for example, contains hundreds of optimization rules that cannot be isolated so where possible it is desirable to not rely on our canonicalization ruleset being complete.

\begin{figure}[ht]
\centering
\begin{subfigure}[b]{0.45\textwidth}
\centering
\begin{lstlisting}[language=Java]
int test1(int a, int b) {
    if (a > b) {
        if (a > b) {
            return 1;
        }
    }
    return 2;
}
\end{lstlisting}
\caption{Unoptimized Java program}
\label{fig:example1unopt}
\end{subfigure}
\hfill
\begin{subfigure}[b]{0.45\textwidth}
\centering
\begin{lstlisting}[language=Java]
int test1(int a, int b) {
    if (a > b) {
        if (true) {
            return 1;
        }
    }
    return 2;
}
\end{lstlisting}
\caption{Optimized Java program after conditional elimination}
\label{fig:example1opt}
\end{subfigure}
\hfill
\begin{subfigure}[b]{0.45\textwidth}
\centering
\begin{lstlisting}[language=Java]
int test1(int a, int b){
    if (a > b) {
        return 1;
    }
    return 2;
}
\end{lstlisting}
\caption{Optimized Java program after conditional elimination and canonicalization}
\label{fig:example1opt2}
\end{subfigure}
\caption{Unoptimized and optimized Java source for a conditional elimination and canonicalization optimization.}
\label{fig:example1}
\end{figure}

\begin{figure}[ht]
\begin{subfigure}[b]{\columnwidth}
\begin{isabelle}
\isamarkuptrue%
\isacommand{definition}\isamarkupfalse%
\ test{\isadigit{1}}{\isacharunderscore}{\kern0pt}initial\ {\isacharcolon}{\kern0pt}{\isacharcolon}{\kern0pt}\ IRGraph\ \isakeyword{where}\isanewline
{\isachardoublequoteopen}test{\isadigit{1}}{\isacharunderscore}{\kern0pt}initial\ {\isacharequal}{\kern0pt}\ irgraph\ {\isacharbrackleft}{\kern0pt}\isanewline
{\isacharparenleft}{\kern0pt}{\isadigit{0}}{\isacharcomma}{\kern0pt}\ {\isacharparenleft}{\kern0pt}StartNode\ {\isacharparenleft}{\kern0pt}Some\ {\isadigit{3}}{\isacharparenright}{\kern0pt}\ {\isadigit{7}}{\isacharparenright}{\kern0pt}{\isacharcomma}{\kern0pt}\ VoidStamp{\isacharparenright}{\kern0pt}{\isacharcomma}{\kern0pt}\isanewline
{\isacharparenleft}{\kern0pt}{\isadigit{1}}{\isacharcomma}{\kern0pt}\ {\isacharparenleft}{\kern0pt}ParameterNode\ {\isadigit{0}}{\isacharparenright}{\kern0pt}{\isacharcomma}{\kern0pt}\ default{\isacharunderscore}{\kern0pt}stamp{\isacharparenright}{\kern0pt}{\isacharcomma}{\kern0pt}\isanewline
{\isacharparenleft}{\kern0pt}{\isadigit{2}}{\isacharcomma}{\kern0pt}\ {\isacharparenleft}{\kern0pt}ParameterNode\ {\isadigit{1}}{\isacharparenright}{\kern0pt}{\isacharcomma}{\kern0pt}\ default{\isacharunderscore}{\kern0pt}stamp{\isacharparenright}{\kern0pt}{\isacharcomma}{\kern0pt}\isanewline
{\isacharparenleft}{\kern0pt}{\isadigit{3}}{\isacharcomma}{\kern0pt}\ {\isacharparenleft}{\kern0pt}FrameState\ {\isacharbrackleft}{\kern0pt}{\isacharbrackright}{\kern0pt}\ None\ None\ None{\isacharparenright}{\kern0pt}{\isacharcomma}{\kern0pt}\ IllegalStamp{\isacharparenright}{\kern0pt}{\isacharcomma}{\kern0pt}\isanewline
{\isacharparenleft}{\kern0pt}{\isadigit{4}}{\isacharcomma}{\kern0pt}\ {\isacharparenleft}{\kern0pt}IntegerLessThanNode\ {\isadigit{2}}\ {\isadigit{1}}{\isacharparenright}{\kern0pt}{\isacharcomma}{\kern0pt}\ VoidStamp{\isacharparenright}{\kern0pt}{\isacharcomma}{\kern0pt}\isanewline
{\isacharparenleft}{\kern0pt}{\isadigit{5}}{\isacharcomma}{\kern0pt}\ {\isacharparenleft}{\kern0pt}BeginNode\ {\isadigit{8}}{\isacharparenright}{\kern0pt}{\isacharcomma}{\kern0pt}\ VoidStamp{\isacharparenright}{\kern0pt}{\isacharcomma}{\kern0pt}\isanewline
{\isacharparenleft}{\kern0pt}{\isadigit{6}}{\isacharcomma}{\kern0pt}\ {\isacharparenleft}{\kern0pt}BeginNode\ {\isadigit{1}}{\isadigit{3}}{\isacharparenright}{\kern0pt}{\isacharcomma}{\kern0pt}\ VoidStamp{\isacharparenright}{\kern0pt}{\isacharcomma}{\kern0pt}\isanewline
{\isacharparenleft}{\kern0pt}{\isadigit{7}}{\isacharcomma}{\kern0pt}\ {\isacharparenleft}{\kern0pt}IfNode\ {\isadigit{4}}\ {\isadigit{6}}\ {\isadigit{5}}{\isacharparenright}{\kern0pt}{\isacharcomma}{\kern0pt}\ VoidStamp{\isacharparenright}{\kern0pt}{\isacharcomma}{\kern0pt}\isanewline
{\isacharparenleft}{\kern0pt}{\isadigit{8}}{\isacharcomma}{\kern0pt}\ {\isacharparenleft}{\kern0pt}EndNode{\isacharparenright}{\kern0pt}{\isacharcomma}{\kern0pt}\ VoidStamp{\isacharparenright}{\kern0pt}{\isacharcomma}{\kern0pt}\isanewline
{\isacharparenleft}{\kern0pt}{\isadigit{9}}{\isacharcomma}{\kern0pt}\ {\isacharparenleft}{\kern0pt}MergeNode\ \ {\isacharbrackleft}{\kern0pt}{\isadigit{8}}{\isacharcomma}{\kern0pt}\ {\isadigit{1}}{\isadigit{0}}{\isacharbrackright}{\kern0pt}\ {\isacharparenleft}{\kern0pt}Some\ {\isadigit{1}}{\isadigit{6}}{\isacharparenright}{\kern0pt}\ {\isadigit{1}}{\isadigit{8}}{\isacharparenright}{\kern0pt}{\isacharcomma}{\kern0pt}\ VoidStamp{\isacharparenright}{\kern0pt}{\isacharcomma}{\kern0pt}\isanewline
{\isacharparenleft}{\kern0pt}{\isadigit{1}}{\isadigit{0}}{\isacharcomma}{\kern0pt}\ {\isacharparenleft}{\kern0pt}EndNode{\isacharparenright}{\kern0pt}{\isacharcomma}{\kern0pt}\ VoidStamp{\isacharparenright}{\kern0pt}{\isacharcomma}{\kern0pt}\isanewline
{\isacharparenleft}{\kern0pt}{\isadigit{1}}{\isadigit{1}}{\isacharcomma}{\kern0pt}\ {\isacharparenleft}{\kern0pt}BeginNode\ {\isadigit{1}}{\isadigit{5}}{\isacharparenright}{\kern0pt}{\isacharcomma}{\kern0pt}\ VoidStamp{\isacharparenright}{\kern0pt}{\isacharcomma}{\kern0pt}\isanewline
{\isacharparenleft}{\kern0pt}{\isadigit{1}}{\isadigit{2}}{\isacharcomma}{\kern0pt}\ {\isacharparenleft}{\kern0pt}BeginNode\ {\isadigit{1}}{\isadigit{0}}{\isacharparenright}{\kern0pt}{\isacharcomma}{\kern0pt}\ VoidStamp{\isacharparenright}{\kern0pt}{\isacharcomma}{\kern0pt}\isanewline
{\highlightnode{{\isacharparenleft}{\kern0pt}{\isadigit{1}}{\isadigit{3}}{\isacharcomma}{\kern0pt}\ {\isacharparenleft}{\kern0pt}IfNode\ {\isadigit{4}}\ {\isadigit{1}}{\isadigit{1}}\ {\isadigit{1}}{\isadigit{2}}{\isacharparenright}{\kern0pt}{\isacharcomma}{\kern0pt}\ VoidStamp{\isacharparenright}{\kern0pt}{\isacharcomma}{\kern0pt}}\isanewline}
{\isacharparenleft}{\kern0pt}{\isadigit{1}}{\isadigit{4}}{\isacharcomma}{\kern0pt}\ {\isacharparenleft}{\kern0pt}ConstantNode\ {\isacharparenleft}{\kern0pt}IntVal\ {\isadigit{3}}{\isadigit{2}}\ {\isacharparenleft}{\kern0pt}{\isadigit{1}}{\isacharparenright}{\kern0pt}{\isacharparenright}{\kern0pt}{\isacharparenright}{\kern0pt}{\isacharcomma}{\kern0pt}\ IntegerStamp\ {\isadigit{3}}{\isadigit{2}}\ {\isacharparenleft}{\kern0pt}{\isadigit{1}}{\isacharparenright}{\kern0pt}\ {\isacharparenleft}{\kern0pt}{\isadigit{1}}{\isacharparenright}{\kern0pt}{\isacharparenright}{\kern0pt}{\isacharcomma}{\kern0pt}\isanewline
{\isacharparenleft}{\kern0pt}{\isadigit{1}}{\isadigit{5}}{\isacharcomma}{\kern0pt}\ {\isacharparenleft}{\kern0pt}ReturnNode\ {\isacharparenleft}{\kern0pt}Some\ {\isadigit{1}}{\isadigit{4}}{\isacharparenright}{\kern0pt}\ None{\isacharparenright}{\kern0pt}{\isacharcomma}{\kern0pt}\ VoidStamp{\isacharparenright}{\kern0pt}{\isacharcomma}{\kern0pt}\isanewline
{\isacharparenleft}{\kern0pt}{\isadigit{1}}{\isadigit{6}}{\isacharcomma}{\kern0pt}\ {\isacharparenleft}{\kern0pt}FrameState\ {\isacharbrackleft}{\kern0pt}{\isacharbrackright}{\kern0pt}\ None\ None\ None{\isacharparenright}{\kern0pt}{\isacharcomma}{\kern0pt}\ IllegalStamp{\isacharparenright}{\kern0pt}{\isacharcomma}{\kern0pt}\isanewline
{\isacharparenleft}{\kern0pt}{\isadigit{1}}{\isadigit{7}}{\isacharcomma}{\kern0pt}\ {\isacharparenleft}{\kern0pt}ConstantNode\ {\isacharparenleft}{\kern0pt}IntVal\ {\isadigit{3}}{\isadigit{2}}\ {\isacharparenleft}{\kern0pt}{\isadigit{2}}{\isacharparenright}{\kern0pt}{\isacharparenright}{\kern0pt}{\isacharparenright}{\kern0pt}{\isacharcomma}{\kern0pt}\ IntegerStamp\ {\isadigit{3}}{\isadigit{2}}\ {\isacharparenleft}{\kern0pt}{\isadigit{2}}{\isacharparenright}{\kern0pt}\ {\isacharparenleft}{\kern0pt}{\isadigit{2}}{\isacharparenright}{\kern0pt}{\isacharparenright}{\kern0pt}{\isacharcomma}{\kern0pt}\isanewline
{\isacharparenleft}{\kern0pt}{\isadigit{1}}{\isadigit{8}}{\isacharcomma}{\kern0pt}\ {\isacharparenleft}{\kern0pt}ReturnNode\ {\isacharparenleft}{\kern0pt}Some\ {\isadigit{1}}{\isadigit{7}}{\isacharparenright}{\kern0pt}\ None{\isacharparenright}{\kern0pt}{\isacharcomma}{\kern0pt}\ VoidStamp{\isacharparenright}{\kern0pt}\isanewline
{\isacharbrackright}{\kern0pt}{\isachardoublequoteclose}%
\end{isabelle}
\caption{Isabelle encoding of the unoptimized Java program from Figure~\ref{fig:example1unopt}}
\end{subfigure}
\hfill

\begin{subfigure}[b]{\columnwidth}
\begin{isabelle}
\isamarkuptrue%
\isacommand{definition}\isamarkupfalse%
\ test{\isadigit{1}}{\isacharunderscore}{\kern0pt}final\ {\isacharcolon}{\kern0pt}{\isacharcolon}{\kern0pt}\ IRGraph\ \isakeyword{where}\ \isanewline
{\isachardoublequoteopen}test{\isadigit{1}}{\isacharunderscore}{\kern0pt}final\ {\isacharequal}{\kern0pt}\ irgraph\ {\isacharbrackleft}{\kern0pt}\isanewline
{\isacharparenleft}{\kern0pt}{\isadigit{0}}{\isacharcomma}{\kern0pt}\ {\isacharparenleft}{\kern0pt}StartNode\ {\isacharparenleft}{\kern0pt}Some\ {\isadigit{3}}{\isacharparenright}{\kern0pt}\ {\isadigit{7}}{\isacharparenright}{\kern0pt}{\isacharcomma}{\kern0pt}\ VoidStamp{\isacharparenright}{\kern0pt}{\isacharcomma}{\kern0pt}\isanewline
{\isacharparenleft}{\kern0pt}{\isadigit{1}}{\isacharcomma}{\kern0pt}\ {\isacharparenleft}{\kern0pt}ParameterNode\ {\isadigit{0}}{\isacharparenright}{\kern0pt}{\isacharcomma}{\kern0pt}\ default{\isacharunderscore}{\kern0pt}stamp{\isacharparenright}{\kern0pt}{\isacharcomma}{\kern0pt}\isanewline
{\isacharparenleft}{\kern0pt}{\isadigit{2}}{\isacharcomma}{\kern0pt}\ {\isacharparenleft}{\kern0pt}ParameterNode\ {\isadigit{1}}{\isacharparenright}{\kern0pt}{\isacharcomma}{\kern0pt}\ default{\isacharunderscore}{\kern0pt}stamp{\isacharparenright}{\kern0pt}{\isacharcomma}{\kern0pt}\isanewline
{\isacharparenleft}{\kern0pt}{\isadigit{3}}{\isacharcomma}{\kern0pt}\ {\isacharparenleft}{\kern0pt}FrameState\ {\isacharbrackleft}{\kern0pt}{\isacharbrackright}{\kern0pt}\ None\ None\ None{\isacharparenright}{\kern0pt}{\isacharcomma}{\kern0pt}\ IllegalStamp{\isacharparenright}{\kern0pt}{\isacharcomma}{\kern0pt}\isanewline
{\isacharparenleft}{\kern0pt}{\isadigit{4}}{\isacharcomma}{\kern0pt}\ {\isacharparenleft}{\kern0pt}IntegerLessThanNode\ {\isadigit{2}}\ {\isadigit{1}}{\isacharparenright}{\kern0pt}{\isacharcomma}{\kern0pt}\ VoidStamp{\isacharparenright}{\kern0pt}{\isacharcomma}{\kern0pt}\isanewline
{\isacharparenleft}{\kern0pt}{\isadigit{5}}{\isacharcomma}{\kern0pt}\ {\isacharparenleft}{\kern0pt}BeginNode\ {\isadigit{8}}{\isacharparenright}{\kern0pt}{\isacharcomma}{\kern0pt}\ VoidStamp{\isacharparenright}{\kern0pt}{\isacharcomma}{\kern0pt}\isanewline
{\isacharparenleft}{\kern0pt}{\isadigit{6}}{\isacharcomma}{\kern0pt}\ {\isacharparenleft}{\kern0pt}BeginNode\ {\isadigit{1}}{\isadigit{3}}{\isacharparenright}{\kern0pt}{\isacharcomma}{\kern0pt}\ VoidStamp{\isacharparenright}{\kern0pt}{\isacharcomma}{\kern0pt}\isanewline
{\isacharparenleft}{\kern0pt}{\isadigit{7}}{\isacharcomma}{\kern0pt}\ {\isacharparenleft}{\kern0pt}IfNode\ {\isadigit{4}}\ {\isadigit{6}}\ {\isadigit{5}}{\isacharparenright}{\kern0pt}{\isacharcomma}{\kern0pt}\ VoidStamp{\isacharparenright}{\kern0pt}{\isacharcomma}{\kern0pt}\isanewline
{\isacharparenleft}{\kern0pt}{\isadigit{8}}{\isacharcomma}{\kern0pt}\ {\isacharparenleft}{\kern0pt}EndNode{\isacharparenright}{\kern0pt}{\isacharcomma}{\kern0pt}\ VoidStamp{\isacharparenright}{\kern0pt}{\isacharcomma}{\kern0pt}\isanewline
{\isacharparenleft}{\kern0pt}{\isadigit{9}}{\isacharcomma}{\kern0pt}\ {\isacharparenleft}{\kern0pt}MergeNode\ \ {\isacharbrackleft}{\kern0pt}{\isadigit{8}}{\isacharcomma}{\kern0pt}\ {\isadigit{1}}{\isadigit{0}}{\isacharbrackright}{\kern0pt}\ {\isacharparenleft}{\kern0pt}Some\ {\isadigit{1}}{\isadigit{6}}{\isacharparenright}{\kern0pt}\ {\isadigit{1}}{\isadigit{8}}{\isacharparenright}{\kern0pt}{\isacharcomma}{\kern0pt}\ VoidStamp{\isacharparenright}{\kern0pt}{\isacharcomma}{\kern0pt}\isanewline
{\isacharparenleft}{\kern0pt}{\isadigit{1}}{\isadigit{0}}{\isacharcomma}{\kern0pt}\ {\isacharparenleft}{\kern0pt}EndNode{\isacharparenright}{\kern0pt}{\isacharcomma}{\kern0pt}\ VoidStamp{\isacharparenright}{\kern0pt}{\isacharcomma}{\kern0pt}\isanewline
{\isacharparenleft}{\kern0pt}{\isadigit{1}}{\isadigit{1}}{\isacharcomma}{\kern0pt}\ {\isacharparenleft}{\kern0pt}BeginNode\ {\isadigit{1}}{\isadigit{5}}{\isacharparenright}{\kern0pt}{\isacharcomma}{\kern0pt}\ VoidStamp{\isacharparenright}{\kern0pt}{\isacharcomma}{\kern0pt}\isanewline
{\isacharparenleft}{\kern0pt}{\isadigit{1}}{\isadigit{2}}{\isacharcomma}{\kern0pt}\ {\isacharparenleft}{\kern0pt}BeginNode\ {\isadigit{1}}{\isadigit{0}}{\isacharparenright}{\kern0pt}{\isacharcomma}{\kern0pt}\ VoidStamp{\isacharparenright}{\kern0pt}{\isacharcomma}{\kern0pt}\isanewline
{\highlightnode{{\isacharparenleft}{\kern0pt}{\isadigit{1}}{\isadigit{3}}{\isacharcomma}{\kern0pt}\ {\isacharparenleft}{\kern0pt}IfNode\ {\isadigit{1}}{\isadigit{9}}\ {\isadigit{1}}{\isadigit{1}}\ {\isadigit{1}}{\isadigit{2}}{\isacharparenright}{\kern0pt}{\isacharcomma}{\kern0pt}\ VoidStamp{\isacharparenright}{\kern0pt}{\isacharcomma}{\kern0pt}}\isanewline}
{\isacharparenleft}{\kern0pt}{\isadigit{1}}{\isadigit{4}}{\isacharcomma}{\kern0pt}\ {\isacharparenleft}{\kern0pt}ConstantNode\ {\isacharparenleft}{\kern0pt}IntVal\ {\isadigit{3}}{\isadigit{2}}\ {\isacharparenleft}{\kern0pt}{\isadigit{1}}{\isacharparenright}{\kern0pt}{\isacharparenright}{\kern0pt}{\isacharparenright}{\kern0pt}{\isacharcomma}{\kern0pt}\ IntegerStamp\ {\isadigit{3}}{\isadigit{2}}\ {\isacharparenleft}{\kern0pt}{\isadigit{1}}{\isacharparenright}{\kern0pt}\ {\isacharparenleft}{\kern0pt}{\isadigit{1}}{\isacharparenright}{\kern0pt}{\isacharparenright}{\kern0pt}{\isacharcomma}{\kern0pt}\isanewline
{\isacharparenleft}{\kern0pt}{\isadigit{1}}{\isadigit{5}}{\isacharcomma}{\kern0pt}\ {\isacharparenleft}{\kern0pt}ReturnNode\ {\isacharparenleft}{\kern0pt}Some\ {\isadigit{1}}{\isadigit{4}}{\isacharparenright}{\kern0pt}\ None{\isacharparenright}{\kern0pt}{\isacharcomma}{\kern0pt}\ VoidStamp{\isacharparenright}{\kern0pt}{\isacharcomma}{\kern0pt}\isanewline
{\isacharparenleft}{\kern0pt}{\isadigit{1}}{\isadigit{6}}{\isacharcomma}{\kern0pt}\ {\isacharparenleft}{\kern0pt}FrameState\ {\isacharbrackleft}{\kern0pt}{\isacharbrackright}{\kern0pt}\ None\ None\ None{\isacharparenright}{\kern0pt}{\isacharcomma}{\kern0pt}\ IllegalStamp{\isacharparenright}{\kern0pt}{\isacharcomma}{\kern0pt}\isanewline
{\isacharparenleft}{\kern0pt}{\isadigit{1}}{\isadigit{7}}{\isacharcomma}{\kern0pt}\ {\isacharparenleft}{\kern0pt}ConstantNode\ {\isacharparenleft}{\kern0pt}IntVal\ {\isadigit{3}}{\isadigit{2}}\ {\isacharparenleft}{\kern0pt}{\isadigit{2}}{\isacharparenright}{\kern0pt}{\isacharparenright}{\kern0pt}{\isacharparenright}{\kern0pt}{\isacharcomma}{\kern0pt}\ IntegerStamp\ {\isadigit{3}}{\isadigit{2}}\ {\isacharparenleft}{\kern0pt}{\isadigit{2}}{\isacharparenright}{\kern0pt}\ {\isacharparenleft}{\kern0pt}{\isadigit{2}}{\isacharparenright}{\kern0pt}{\isacharparenright}{\kern0pt}{\isacharcomma}{\kern0pt}\isanewline
{\isacharparenleft}{\kern0pt}{\isadigit{1}}{\isadigit{8}}{\isacharcomma}{\kern0pt}\ {\isacharparenleft}{\kern0pt}ReturnNode\ {\isacharparenleft}{\kern0pt}Some\ {\isadigit{1}}{\isadigit{7}}{\isacharparenright}{\kern0pt}\ None{\isacharparenright}{\kern0pt}{\isacharcomma}{\kern0pt}\ VoidStamp{\isacharparenright}{\kern0pt}{\isacharcomma}{\kern0pt}\isanewline
{\highlightnode{{\isacharparenleft}{\kern0pt}{\isadigit{1}}{\isadigit{9}}{\isacharcomma}{\kern0pt}\ {\isacharparenleft}{\kern0pt}ConstantNode\ {\isacharparenleft}{\kern0pt}IntVal\ {\isadigit{1}}\ {\isacharparenleft}{\kern0pt}{\isadigit{1}}{\isacharparenright}{\kern0pt}{\isacharparenright}{\kern0pt}{\isacharparenright}{\kern0pt}{\isacharcomma}{\kern0pt}\ VoidStamp{\isacharparenright}{\kern0pt}}\isanewline}
{\isacharbrackright}{\kern0pt}{\isachardoublequoteclose}%
\end{isabelle}
\caption{Isabelle encoding of the optimized Java program from Figure~\ref{fig:example1opt}}
\end{subfigure}
\caption{Unoptimized and optimized Isabelle/HOL encodings of the Java programs in Figure~\ref{fig:example1}.}
\label{fig:example1enc}
\end{figure}

\begin{figure}[ht]
\begin{isabelle}
\isamarkuptrue%
\isacommand{corollary}\isamarkupfalse%
\ {\isachardoublequoteopen}{\isacharparenleft}{\kern0pt}runConditionalElimination\ test{\isadigit{1}}{\isacharunderscore}{\kern0pt}initial{\isacharparenright}{\kern0pt}\ {\isasymapprox}\isactrlsub s\ test{\isadigit{1}}{\isacharunderscore}{\kern0pt}final{\isachardoublequoteclose}\isanewline
\ \ %
\isacommand{by}\isamarkupfalse%
\ eval%
\end{isabelle}
\caption{Comparing the GraalVM optimized program to the Isabelle/HOL optimized program.}
\label{fig:comparison}
\end{figure}

As with validation of the IR semantics,
we utilize our translator from the GraalVM IR to an encoding of the IR in our Isabelle/HOL representation.
We use the translator to export an Isabelle/HOL encoded IR of the unoptimized and optimized program.
For example,
Figure \ref{fig:example1enc} represents the test programs from Figure \ref{fig:example1} in Isabelle/HOL.
In this example,
the only change between the programs is the addition a new constant node, \textsl{node 19},
which is the constant \textsf{true} (represented by 1) and the replacement of the reference to node 4 in the \Node{If} with id 13
with a reference to the new node 19.

The optimization specification in Isabelle/HOL is applied to the encoded unoptimized graph.
This leaves two versions of the optimized program,
one that was optimized by GraalVM and encoded,
and one that was optimized by our Isabelle specification.
These programs are then compared to determine whether they are structurally equivalent.
We can be reasonably confident in our optimization specification if all such programs are structurally equivalent.
Figure \ref{fig:comparison} demonstrates this final step for our initial example.
\texttt{runConditionalElimination} is applied to the unoptimized encoded program to produce an Isabelle optimized program.
The optimized program is then structurally compared via $\approx_s$ to the optimized encoded program.

When developing the specification of optimizations within the conditional elimination phase,
this validation approach was utilized.
Our approach was to develop a specification alongside validating tests which roughly follows a test-driven development methodology.
This was useful in increasing confidence that the specification was accurate during development,
particularly as the code that performs optimizations can often be nested within code that is irrelevant for the optimization rule specification in Isabelle, 
such as analytics and debugging, so elements of the transformation can be easily missed.

Despite this approach,
a number of bugs were identified after generating additional test cases:
\begin{description}
    \item[Excessively optimizing branches:]
        The conditional elimination phase is only responsible for replacing branch conditions that are always true or always false with the constants true or false.
        Due to initially including a canonicalization pass in the generated test cases,
        the specification mistakenly also replaced the redundant branches with their appropriate successors.
        This approach was not faithful to the multi-pass approach of the compiler.
    \item[Negated conditions:]
        The conditional elimination phase constructs a stack of conditions from each basic block it enters, popping the stack when exiting.
        When traversing the false branch,
        the negated condition is pushed to the stack.
        This was not initially accounted for in all optimizations defined in the phase resulting in missed optimizations identified by the tests.
    \item[Control-flow traversal:]
        During traversal of the graph,
        type information is pushed onto a stack when a basic block is entered, identified by a \texttt{BeginNode},
        and popped when left, identified by an \texttt{EndNode}.
        Through testing we identified that this was an invalid approach as our assumed invariant of balanced begin and end nodes did not hold for the GraalVM IR.
        When there are successive end nodes with no intervening nodes,
        they are collapsed into a single end node.
        This necessitates a more complicated dominator based traversal as implemented in the compiler.
\end{description}

The conditional elimination phase makes use of two sources of information to determine 
if a condition in an \Node{If} can be replaced by either \textsf{true} or by \textsf{false}:
\begin{itemize}
\item
a stack of conditions collected from \Node{If}s on the path from the start node that are either known to be true or known to be false, and
\item
a stack of stamps for nodes, each containing lower and upper bounds on their values, also collected on the path.
\end{itemize}
Our Isabelle specification of refining the stamp based on the path,
did not initially match that within the GraalVM due to an off-by-one error
which meant that our Isabelle specification could miss optimizations performed by the compiler.
This issue was picked up by inspection of the Isabelle specification 
but was not picked up by the testing because in all of the test cases in the compiler,
the optimization was recognized using the stack of conditions and hence the faulty check against the stamp was masked.
This issue indicated that more complex test cases were necessary to fully test this aspect of the optimization.
When such a test case was developed, we discovered that our corrected Isabelle performed an optimization whereas the GraalVM did not.
This was because our Isabelle updated the stamp from the top of the stack of stamps, whereas GraalVM updated from the original stamp each time.

\section{Related Work}

Since its introduction by McKeeman \cite{McKeeman1998DifferentialTesting}, differential testing has been widely used for checking the consistency of two (or more) alternative implementations of a common specification, such as two different C compilers.
The basic idea is to generate many example inputs, feed each input through both implementations, and then check that the outputs are equivalent.  This is similar to fuzzing \cite{fuzz} but with a much stronger oracle for the correctness of each test because comparing the two outputs means that we are checking the functional behavior of the two implementations (checking one implementation against the other), whereas 
fuzzing typically just detects crashes and runtime errors in a single implementation.  

For example, Chen \emph{et al.}~\cite{Chen2016DifferentialTesting_PLDI} used differential testing to compare two JVM implementations, using Markov chain Monte Carlo sampling to generate new Java input programs by mutating a pool of given programs.

Similarly, Marmsoler and Brucker \cite{Marmsoler2022ConformanceTesting} used differential testing to check the conformance of their formal Isabelle/HOL semantics for Solidity against the reference implementation of the Solidity compiler with execution of the generated bytecode on the Ethereum Blockchain.  They used grammar-based fuzzing to generate the input Solidity programs, using a restricted grammar to ensure that the generated programs were type-correct. 
This is similar to our approach to testing our Isabelle/HOL formal semantics of the GraalVM IR in Section \ref{sec:execsemantics}, except that as well as generating some very simple input Java methods (to test integer operators, Section \ref{sec:arith}) we also extracted Java methods from the existing GraalVM test suite of thousands of hand-written tests.  Those unit tests are carefully designed to exercise most of the different semantic behaviors of Java, so reusing these tests gives us good coverage of various Java features.

Another common use of differential testing is to test the optimization features of a compiler, by turning on and off various optimization levels and comparing the non-optimizing and optimizing versions of the compiler against each other.  For example, Chen \emph{et al.}~\cite{chen:opt-exploration22} found 17 bugs in GCC and LLVM by using machine-learning to learn relationships between test programs and optimization settings, and then using differential testing to find incorrect optimizations with various optimization settings.
This is somewhat similar to our differential testing in Section \ref{sec:iropt}, where we are validating our Isabelle/HOL definitions of optimization phases against the GraalVM compiler, by capturing the IR graph within the compiler before and after the optimization phase, then using differential testing to check that our Isabelle/HOL transformations give an equivalent result to the GraalVM compiler.

\section{Conclusions} \label{sec:conclusions}

We have shown that differential testing is a powerful technique for finding differences between two semantics.  We have used it to find differences between:
\begin{enumerate}
  \item our formal Isabelle models of GraalVM IR semantics and the machine code generated by the GraalVM compiler; and
  \item our Isabelle models of IR optimization transformations and the IR optimizations performed within the GraalVM compiler.
\end{enumerate}

Our approach to validating our Isabelle/HOL models has leveraged the extensive existing test suites of the GraalVM compiler.  Apart from our generation of boundary tests for the integer arithmetic operators, we have relied completely on the existing GraalVM compiler unit tests for the inputs to our differential testing.

In the future, we plan to extend our differential testing further by \emph{generating} input Java methods 
using grammar-based fuzzing approaches, similar to Marmsoler and Brucker's approach for Solidity \cite{Marmsoler2022ConformanceTesting}.  
In addition, we plan to add hooks into the GraalVM compiler optimization phases to dump the IR graphs before and after individual optimization transformations, rather than before and after an entire optimization phase.  This will enable us to test each individual optimization transformation in isolation, more thoroughly, rather than having to test a whole optimization phase at once.
This should further strengthen our validation of the optimization transformations we have defined in Isabelle/HOL.

\begin{acks}
Mark Utting's position and Brae Webb's scholarship are both funded in part by a gift from Oracle Labs.
Thanks especially to Cristina Cifuentes, Paddy Krishnan and Andrew Craik from Oracle Labs Brisbane for their helpful feedback, and to Gerg\"o Barany and the rest of the Oracle GraalVM compiler team for answering questions.  
Thanks also to Ethan Bulmer for his honours thesis work on translating IR graphs into Isabelle syntax. 
\end{acks}

\bibliographystyle{ACM-Reference-Format}
\bibliography{references,veriopt}

\end{document}